# Direct measurement of proximity-induced magnetism at the buried interface between a topological insulator and a ferromagnet

Changmin Lee[1], Ferhat Katmis[1,2], Pablo Jarillo-Herrero[1], Jagadeesh S. Moodera[1,2] and Nuh Gedik[1*]

[1] Department of Physics, Massachusetts Institute of Technology, Cambridge MA 02139, USA.

[2] Francis Bitter Magnet Laboratory & Plasma Science and Fusion Center, Massachusetts Institute of Technology, Cambridge, MA 02139, USA.

* To whom correspondence and requests for materials should be addressed. Email: gedik@mit.edu

**When a topological insulator (TI) is in contact with a ferromagnet, both time reversal and inversion symmetries are broken at the interface[1,2]. An energy gap is formed at the TI surface, and its electrons gain a net magnetic moment through short-range exchange interactions. Magnetic TIs can host various exotic quantum phenomena, such as massive Dirac fermions[3], Majorana fermions[4], the quantum anomalous Hall effect[5,6] and chiral edge currents along the domain boundaries[1,2]. However, selective measurement of induced magnetism at the buried interface has remained a challenge. Using magnetic second harmonic generation, we directly probe both the in-plane and out-of-plane magnetizations induced at the interface between the ferromagnetic insulator (FMI) EuS and the three-dimensional TI $Bi_2Se_3$. Our findings not only allow characterizing magnetism at the TI-FMI interface but also lay the groundwork for imaging magnetic domains and domain boundaries at the magnetic TI surfaces.**

A topological insulator (TI) is a bulk electrical insulator with a two-dimensional metallic surface. The linearly dispersing surface states remain gapless and the spin and momentum channels of the surface electrons are locked to each other as long as time-reversal symmetry (TRS) is preserved[1,2]. When TRS is externally broken with moderate perturbations, however, a TI does not simply transform into a trivial insulator. While an energy gap opens up and the spin degeneracy is lifted at the Dirac point[3], the surface electrons acquire a nontrivial spin texture distinct from that of a TI or a trivial insulator[7]. TRS can be broken on a TI surface either by introducing magnetic dopants, or by directly growing a ferromagnetic insulator (FMI) on top. While most of the past research has been focused on the former approach[3,7-10], the TI-FMI interfaces provide a cleaner method by inducing a more uniform long-range magnetization across the TI surface. Moreover, such interfaces do not introduce impurity scattering centres to the surface electrons.

Recently, the EuS-$Bi_2Se_3$ interface has emerged as a novel TRS-broken TI system[11,12]. By itself, EuS is an electrical insulator (energy gap = 1.65 eV) that becomes ferromagnetic below ~17 K (= $T_{Curie}$), and its spins are aligned parallel to the basal plane[13]. However, when EuS is in close proximity to $Bi_2Se_3$, electrons at the EuS-$Bi_2Se_3$ interface acquire a canted magnetic moment, causing TRS to be broken at the TI surface. So far, conventional magnetometry and linear magneto-optical Kerr effects have been used to measure the magnetic properties of these systems[11,12], but these measurements were dominated by the contribution from the bulk magnetism of the EuS film. Furthermore, the nature of the proximity-induced magnetism at the TI-FMI interface is not completely understood.

Magnetic second harmonic generation (MSHG) is a powerful optical tool capable of selectively probing interface magnetism[14], as symmetry arguments restrict the MSHG signal to be generated only from magnetized surfaces and interfaces of a centrosymmetric material. MSHG has been successfully applied to measure magnetism either at the Fe surface[15], or Co/Cu[16] and $SrTiO_3$/$La_{1-x}Sr_xMnO_3$ interfaces[17] to name a few. In this work, we reveal both the magnetic and crystal symmetries at the interface of $EuS$-$Bi_2Se_3$ heterostructures with MSHG. We were able to simultaneously measure the in-plane and out-of-plane components of the proximity-induced ferromagnetism at the buried $EuS$-$Bi_2Se_3$ interface.

The MSHG experiment was carried out with femtosecond laser pulses from an amplified Ti:sapphire system in a transmission geometry (Fig. 1a). A ring-shaped neodymium magnet was used to produce either an in-plane (up to 300 Oe) or an out-of-plane magnetic field (up to 4000 Oe) at the sample, and a set of a half-wave plate and a polarizer was placed to control the input and output light polarizations, respectively. Fluorescent filters were placed before the photomultiplier tubes (PMTs) to prevent any residual fundamental frequency laser pulses from being detected. The experiment was performed under two different configurations: (i) two PMTs were simultaneously used to measure the SHG Faraday rotation (Fig. 1a) or (ii) a single PMT was used to measure the SHG intensity as a function of the input polarization angle (MSHG rotational anisotropy (MSHG-RA) patterns[18] shown in Fig. 2).

Under the first experimental configuration (Fig. 1a), SHG Faraday rotation angles were measured from a 7 nm-7 QL (EuS and $Bi_2Se_3$ thicknesses, respectively; QL: quintuple layer) hybrid heterostructure sample. When a magnetic field (300 Oe) is applied to the sample along the in-plane direction, magnetization at the interface causes the SHG polarization plane to be rotated. When the direction of the magnetic field is reversed, the polarization plane is rotated in the opposite direction. The amount of this polarization rotation is defined as the SHG Faraday rotation. In Figs. 1b,c, it is clearly seen that a large SHG Faraday rotation sets in slightly below 17 K ($T_{Curie}$), verifying its magnetic origin. It is also important to note that SHG Faraday rotation is allowed in samples with in-plane magnetization at normal incidence, unlike the linear Faraday effect that is only sensitive to out-of-plane magnetization[14].

Since there are four distinct interfaces in a EuS-$Bi_2Se_3$ heterostructure film (Fig. 1d), it is important to identify the interfaces contributing to SHG. We thus prepared both (i) a bare sapphire (0001) substrate and (ii) a EuS film (5 nm thick) grown on sapphire (0001) as control samples. We could not detect any measurable SHG signal from either sample down to 4 K. The same observation holds even when a magnetic field was applied either in the in-plane (up to 300 Oe) or the out-of-plane (up to 4000 Oe) direction. Therefore, we conclude that all surface dipole SHG contributions of a EuS-$Bi_2Se_3$ heterostructure arise only from the EuS-$Bi_2Se_3$ and $Bi_2Se_3$-sapphire interfaces, and not from the top EuS surface or the bottom sapphire surface (Fig. 1d). We also note that since SHG is not generated from the bulk of the magnetized EuS film, SHG Faraday rotation arises from the proximity-induced ferromagnetism at the top surface of the $Bi_2Se_3$ film.

In order to quantitatively characterize the strength of interface ferromagnetism and reveal the magnetic symmetry of EuS-Bi$_2$Se$_3$ heterostructures, we took MSHG-RA measurements from the same (7 nm-7 QL) heterostructure sample. SHG intensity was measured as a function of input polarization angle, while an in-plane (Fig. 2a) or an out-of-plane (Fig. 2d) magnetic field was applied to the sample. The output polarization of SHG was selected to be either parallel (PA) or perpendicular (CR, or crossed) to the input polarization. At 295 K, no discernable change in the RA patterns was measured in the presence of a magnetic field. However, at 4 K (<T$_{Curie}$), a large difference in the MSHG-RA pattern was observed between +300 Oe and -300 Oe of in-plane magnetic fields applied to the sample (Figs. 2b,c).

The MSHG-RA patterns can be divided into contributions from i) the nonmagnetic crystal and ii) the interface ferromagnetism[14,19].

$$E(2\omega) \propto P(2\omega) = P_{cr}(2\omega) + P_{mag}(2\omega)$$
$$= \chi^{(2)}_{cr} E(\omega)E(\omega) + \chi^{(2)}_{mag}(M)E(\omega)E(\omega), \quad (1)$$

where $P(2\omega)$ is the second-order electric polarization, $\chi^{(2)}$ is the second-order electric susceptibility tensor, $M$ is magnetization, and $E(\omega)$ is the electric field of the incident light of frequency $\omega$. The tensor $\chi^{(2)}_{mag}(M)$ obtains a nonzero value only below T$_{Curie}$. Following Refs. 19 and 20, $E(2\omega)$ can be written as a function of the input polarization angle $\phi$, the thin film orientation angle $\phi_1$, and the direction of the magnetic field $\phi_2$:

$$E(2\omega) = E_{cr}(2\omega) + E_{mag}(2\omega)$$

$$= A\cos(3\phi + \phi_1) \pm B(M)\cos(\phi + \phi_2), \qquad (2)$$

where $A\cos(3\phi + \phi_1)$ term corresponds to the $3m$ symmetry of the Bi$_2$Se$_3$ surface, and $B(M)\cos(\phi + \phi_2)$ term reflects the strength and direction of the in-plane magnetization at the EuS-Bi$_2$Se$_3$ interface. The latter term is odd (changes sign) with respect to magnetization. Since MSHG intensity, not the electric field strength, is measured in the actual experiment, the MSHG-RA signal consists of the following three terms:

$$I(2\omega) = A^2\cos^2(3\phi + \phi_1) \pm 2C(M)\cos(3\phi + \phi_1)\cos(\phi + \phi_2)$$

$$+ B^2(M^2)\cos^2(\phi + \phi_2). \qquad (3)$$

The first SHG term containing $A^2$ purely comes from the crystal, and the third term containing $B^2(M^2)$ is from the interface magnetism. The second term is caused by the interference of the crystalline and magnetic SHG terms. The interference term $C(M)$ is odd with respect to the magnetization of the sample, and thus gives rise to the large nonlinear Faraday rotation in Figs. 1b,c and the differential MSHG-RA patterns in Figs. 2b,e. The MSHG-RA data due to in-plane magnetization are fitted remarkably well to Eq. (3), as shown by the orange and purple dots (data) and lines (fits) in Figs. 2b,e.

We now turn to the effect of out-of-plane magnetization on the MSHG-RA patterns. Out-of-plane magnetic moments are more important as they are responsible for breaking TRS in a TI system. While in-plane magnetization is easily induced even with a few tens of Oersteds of magnetic field, out-of-plane magnetization requires a stronger magnetic field (up to 3 T for complete saturation)[11]. Since it is difficult to align the out-of-plane magnetic domains without affecting the in-plane domains, we applied a tilted (~ 4°) magnetic field to the sample so that both the in-plane and the out-of-plane magnetizations are induced along a preferred direction at the interface. As described in Ref. 19, out-of-plane magnetization causes the MSHG-RA patterns to be rotated:

$$I(2\omega) = A^2 \cos^2(3\phi + \phi_1 \pm \phi(M_\perp))$$
$$\pm 2C(M_\parallel)\cos(3\phi + \phi_1 \pm \phi(M_\perp))\cos(\phi + \phi_2)$$
$$+ B^2(M_\parallel^2)\cos^2(\phi + \phi_2), \qquad (4)$$

where $M_\parallel$ and $M_\perp$ are in-plane and out-of-plane magnetizations, respectively, and $\phi(M_\perp)$ is the rotation angle of the RA pattern due to out-of-plane magnetization. Figs. 2e,f show the MSHG-RA patterns when a tilted out-of-plane magnetic field of 4000 Oe is applied toward (orange) or away from (purple) the sample. A small rotation of the MSHG-RA pattern was observed, indicating the presence of a net out-of-plane magnetic moment at the EuS-Bi$_2$Se$_3$ interface. We note that the magnitude of polarization rotation angle ($\sim 1°$) measured in this MSHG-RA pattern is orders of magnitude larger than that reported from a previous linear magneto-optic Kerr effect measurements ($\sim 100$ μrad)[12].

To further verify that the MSHG signal comes from the EuS-Bi$_2$Se$_3$ interface, and not from the magnetic bulk of the EuS film, we took MSHG-RA measurements from samples of different EuS thicknesses. While the Bi$_2$Se$_3$ thickness was fixed at 7 QL, the EuS thickness was varied from 2 to 10 nm (Fig. 3a). In Figs. 3b,c, values of $C(M)/A^2$ and $C(M)/A$ (normalized) are plotted as a function of EuS thickness. Since $C(M) \propto \chi_{cr}^{(2)} \cdot \chi_{mag}^{(2)}$ and $A \propto \chi_{cr}^{(2)}$, it follows that $C(M)/A^2 \propto \chi_{mag}^{(2)}/\chi_{cr}^{(2)}$, and $C(M)/A \propto \chi_{mag}^{(2)}$. For both quantities, two sets of data can be acquired for the PA and CR polarization setups. If the magnetic signal were to come from the bulk of EuS film, one would expect a monotonically increasing behavior in $C(M)/A^2$ and $C(M)/A$, which is not the case in Figs 3b,c. A striking feature here is the large variation of the magnetic signal across different EuS thicknesses.

There are two possible explanations for the fluctuations in the magnetic signal. First, since MSHG is interface sensitive, $C(M)/A^2$ and $C(M)/A$ values may simply correspond to the different interface quality of each sample. In this case, however, magnetic signal for the PA and CR polarization setups should increase or decrease together, unlike Figs. 3b,c where the PA and CR magnetic signals often follow an opposite trend with respect to each other. Another possibility is the existence of spin-polarized states in the EuS-Bi$_2$Se$_3$ heterostructures similar to those observed in the Co/Cu films[21] and Au/Co/Au trilayers[22]. When the thickness of a material becomes finite, its electrons can form a quantum well state bound inside the film. Moreover, in a ferromagnetic system, the majority and the minority spin states can become relatively more or less bound depending on the strength of the potential barrier exerted by the neighboring material. Such spin-polarized quantum well states can give rise to an oscillating behavior of the magnetic signal as a function of sample thickness[21,22] (see Supplementary Information). A detailed band structure calculations of the EuS-Bi$_2$Se$_3$ heterostructures can help to answer this question[23].

The out-of-plane magnetic signal is more uniform across samples of different EuS thicknesses (Fig. 3d), compared to the in-plane response. It is important to note that all five samples exhibit both in-plane and out-of-plane magnetizations, and that the magnetic signal does not uniformly increase with EuS thickness, indicating the strong interface sensitivity of MSHG.

In order to rule out the possibility that MSHG probes bulk magnetism of Bi$_2$Se$_3$ films, we now proceed to measurements taken on samples of different Bi$_2$Se$_3$ thicknesses. While EuS thickness was fixed at 5 nm, Bi$_2$Se$_3$ thickness was varied from 1 QL to 10 QL (Fig. 4a). As shown in Fig. 4b, $C(M)/A^2 (\propto \chi^{(2)}_{mag}/\chi^{(2)}_{cr})$ sharply decreases with Bi$_2$Se$_3$ thickness, which is due to the increasing bulk nonmagnetic SHG contributions from thicker samples. It has been reported from earlier studies on Bi$_2$Se$_3$ single crystals that SHG contains bulk contributions[24,25] due to band bending effects near the surface. A similar band bending effect in Bi$_2$Se$_3$ films induced by the SiC substrate of Bi$_2$Se$_3$ films was also reported from a previous ARPES measurement[26]. When $C(M)/A (\propto \chi^{(2)}_{mag})$ is plotted against Bi$_2$Se$_3$ thickness (Fig. 4c), we still observe a decreasing trend. Such behavior is attributed to the increasing absorption of MSHG from thicker Bi$_2$Se$_3$ samples (penetration depth for a 400 nm laser beam is approximately 10 nm, or 10 QL[27]), and indicates that MSHG is not generated from the bulk of the Bi$_2$Se$_3$ film. Out-of-plane response is more uniform across samples of different Bi$_2$Se$_3$ thicknesses (Fig. 4d), except for the 1 QL sample that barely exhibits a nonzero value.

We measured both the in-plane and out-of-plane ferromagnetism induced at the EuS-Bi$_2$Se$_3$ interface due to proximity effect, and demonstrated the interface sensitivity of MSHG. The capability of reducing band-bending effects and tuning chemical potential near the Dirac point (e.g. through electrostatic gating)[6,10,28] should enhance the effectiveness of MSHG as a tool of probing the buried interface that is not easily accessed by other conventional techniques.

Having established that MSHG is an effective tool of probing the magnetic symmetry of TI-FMI interfaces, it can be extended to imaging magnetic domains and domain boundaries at the interface[29]. One particular interest in the study of magnetic TIs is the existence of chiral edge states along the ferromagnetic domain boundaries[1,2]. Such dissipationless chiral currents can be potentially used for realizing topological magnetoelectric effects and novel devices based on spintronics. MSHG imaging can be a powerful tool of studying such chiral modes, as it can potentially distinguish different types of magnetic domain walls (Bloch or Néel)[14] or visualize spin texture at the domain boundary.

**Methods**

EuS-$Bi_2Se_3$ heterostructure samples were grown using a molecular beam epitaxy (MBE) system, as described in Ref. 11. The top EuS surface was capped with 5 nm of amorphous $Al_2O_3$ in order to protect the volatile EuS surface. All MSHG measurements were carried out using a Ti:sapphire laser amplifier system operating at 785 nm with a pulse duration of 50 fs. In all measurements, the incident laser pulse was focused onto a 50 μm spot at the sample using a standard plano-convex lens, and the fluence of the incident laser excitation did not exceed 3 mJ cm$^{-2}$, which is below the damage threshold of all heterostructure samples.

**Acknowledgements**

The authors acknowledge technical assistance from and helpful discussions with Z. Alpichshev and F. Mahmood. C.L acknowledges a Samsung Scholarship from the Samsung Foundation of Culture. This work is supported by the STC Center for Integrated Quantum Materials under NSF grant DMR-1231319 (optical setup, data acquisition, and data analysis), the Division of Materials Research under NSF grant DMR-1207469 (maintenance of the MBE system), the MIT CMSE IRG of the MRSEC Program under NSF grant DMR-0819762 (upgrade of the MBE system), the DOE, Basic Energy Science Office, Division of Material Sciences and Engineering under award DE-SC0006418 (growth of the thin film heterostructures) and the ONR grant N00014-13-1-0301 (characterization of the thin film heterostructures).

**Author contributions**

C.L. performed the experiments, analyzed the data and wrote the initial draft of the manuscript. F.K. prepared the thin film heterostructure samples. C.L., F.K. and N.G. conceived the project. All authors contributed to the understanding of the data and editing the manuscript. P.J.H. and J.S.M. supervised the thin film heterostructure growth and N.G. supervised the optical measurements.

**Competing financial interests**

The authors declare no competing financial interests.

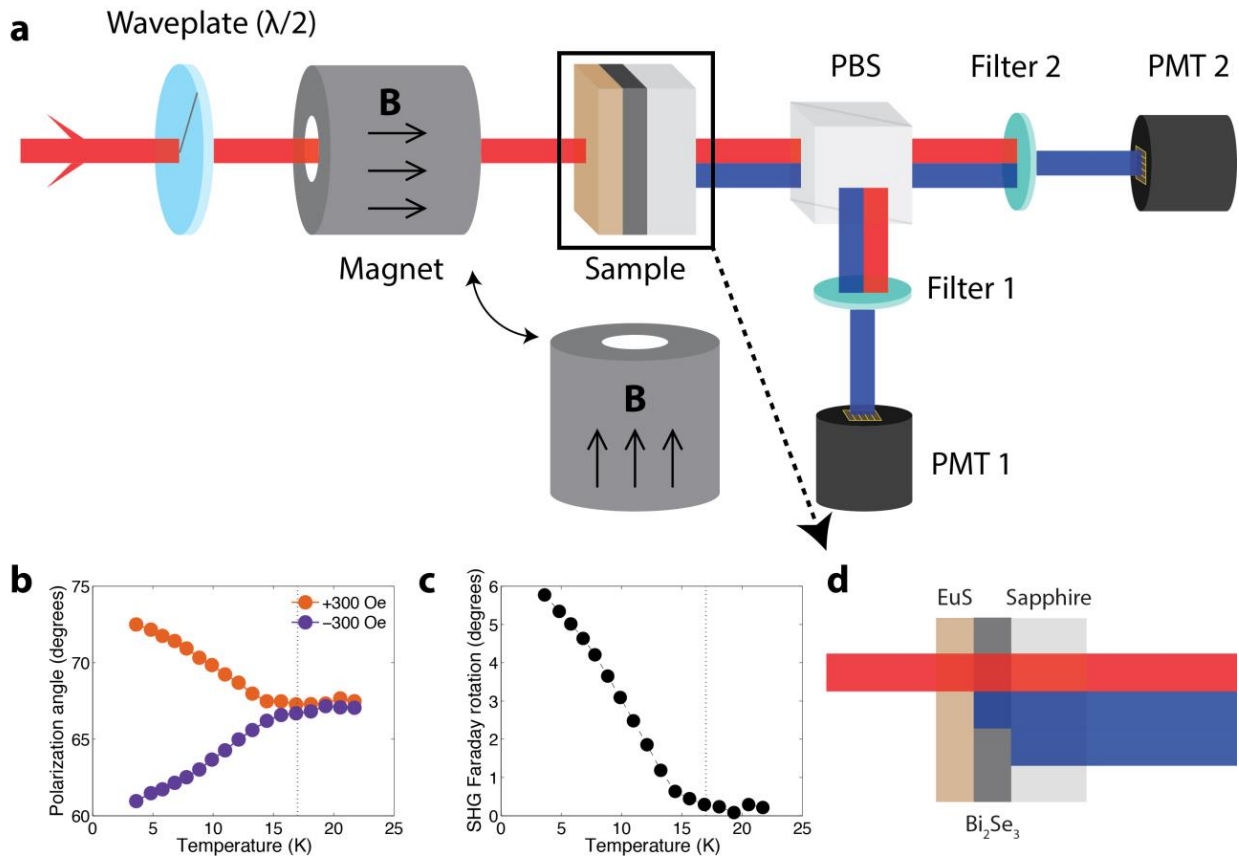

**Figure 1 | SHG Faraday effect measurement. a**, Experimental setup. PBS: polarizing beam splitter, PMT: photomultiplier tube. Rotation of the SHG polarization plane was calculated by using two PMTs that separately measure SHG intensities of *s* and *p* polarizations, respectively. Either an in-plane or an out-of-plane magnetic field was applied to the sample. **b**, Output SHG polarization angle and **c**, SHG Faraday rotation angle plotted against temperature for the 7 nm (EuS) – 7 QL ($Bi_2Se_3$) sample. **d**, In EuS-$Bi_2Se_3$ heterostructures, surface SHG is generated from the EuS-$Bi_2Se_3$ and $Bi_2Se_3$-sapphire interfaces.

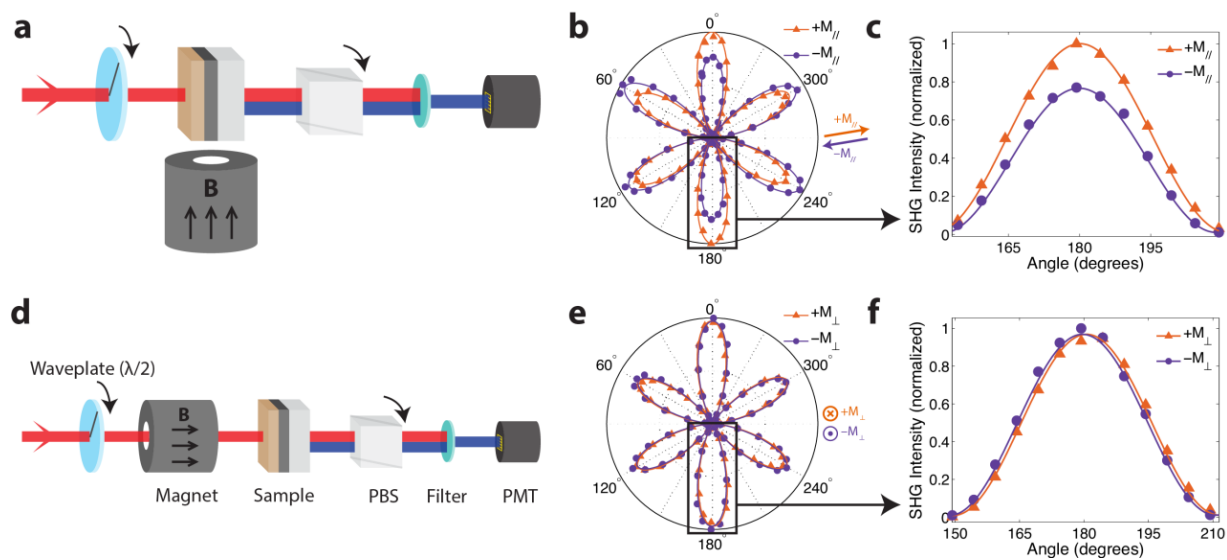

**Figure 2 | MSHG-RA patterns of the EuS-Bi$_2$Se$_3$ heterostructures. a,d**, Experimental setup of the MSHG RA measurements. Magnitude of SHG was measured as a function of input polarization angle while an **a** in-plane or an **d** out-of-plane magnetic field was applied to the sample. A set of a half-wave plate and a polarizer was rotated simultaneously so that the output SHG polarization was set to be either parallel or perpendicular to the input polarization. Only the parallel polarization measurements are shown here. **b**, SHG Intensity as a function of input polarization angle from a 7 nm (EuS)-7 QL (Bi$_2$Se$_3$) sample. In-plane magnetic fields of +300 Oe (orange) and -300 Oe (purple) were applied to the sample at 4 K. **c,** An enlarged plot near 180°. **e,** Same as **b**, but out-of-plane magnetic fields of +4000 Oe (orange) and -4000 Oe (purple) were applied to the sample. A small rotation of the MSHG-RA pattern is observed in **f**, which is an enlarged plot of the squared area in **e**. In **b,e,** magnetic field directions are denoted by orange and purple arrows, dots and crosses.

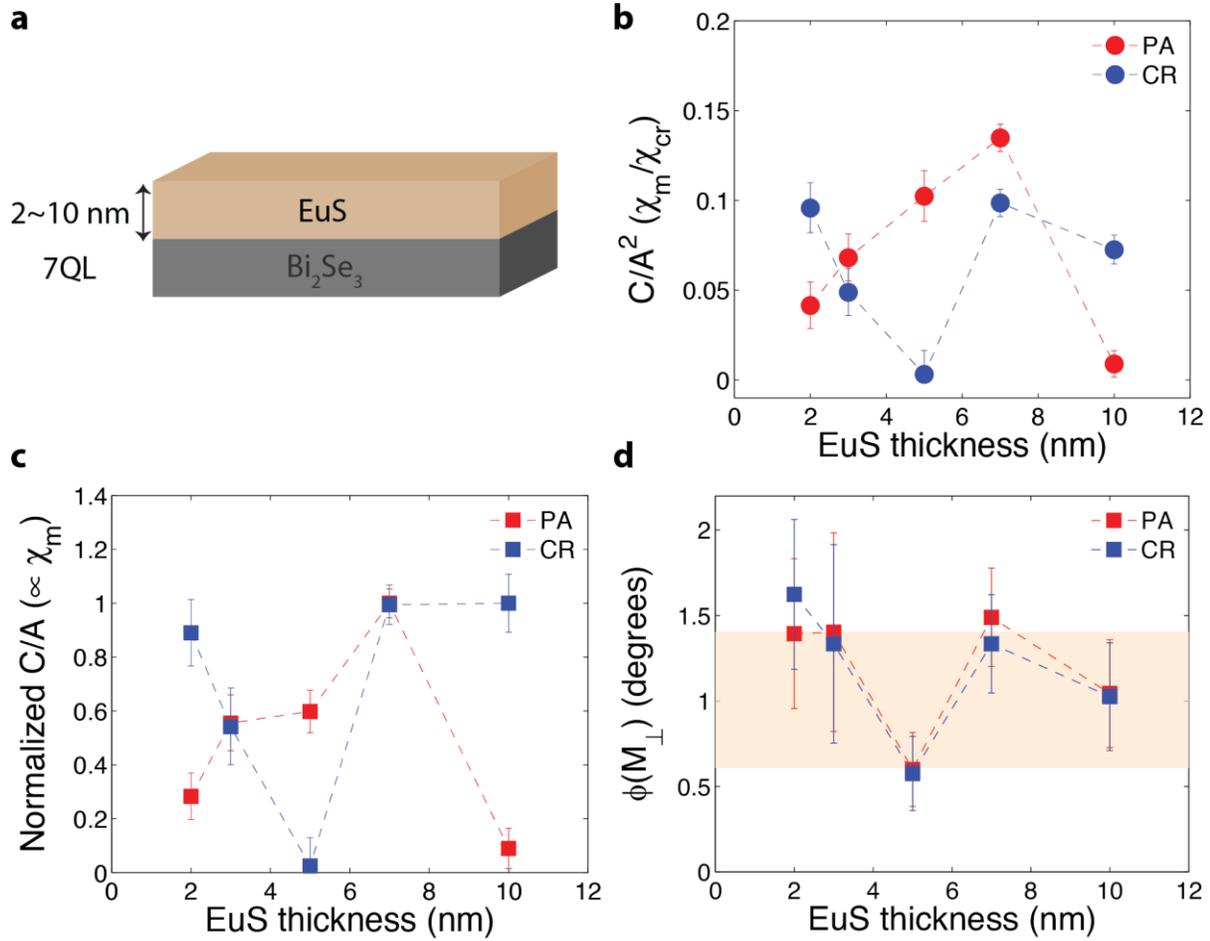

**Figure 3 | EuS thickness dependence. a**, MSHG parameters were measured from samples of various EuS thicknesses (2-10 nm). Bi$_2$Se$_3$ thickness is fixed at 7 QL. **b**, $C(M)/A^2$ and **c**, $C(M)/A$ are plotted as a function of EuS thickness. Both values exhibit a strong variance over different EuS thicknesses. **d**, Out-of-plane response, $\phi(M_\perp)$, for different EuS thicknesses.

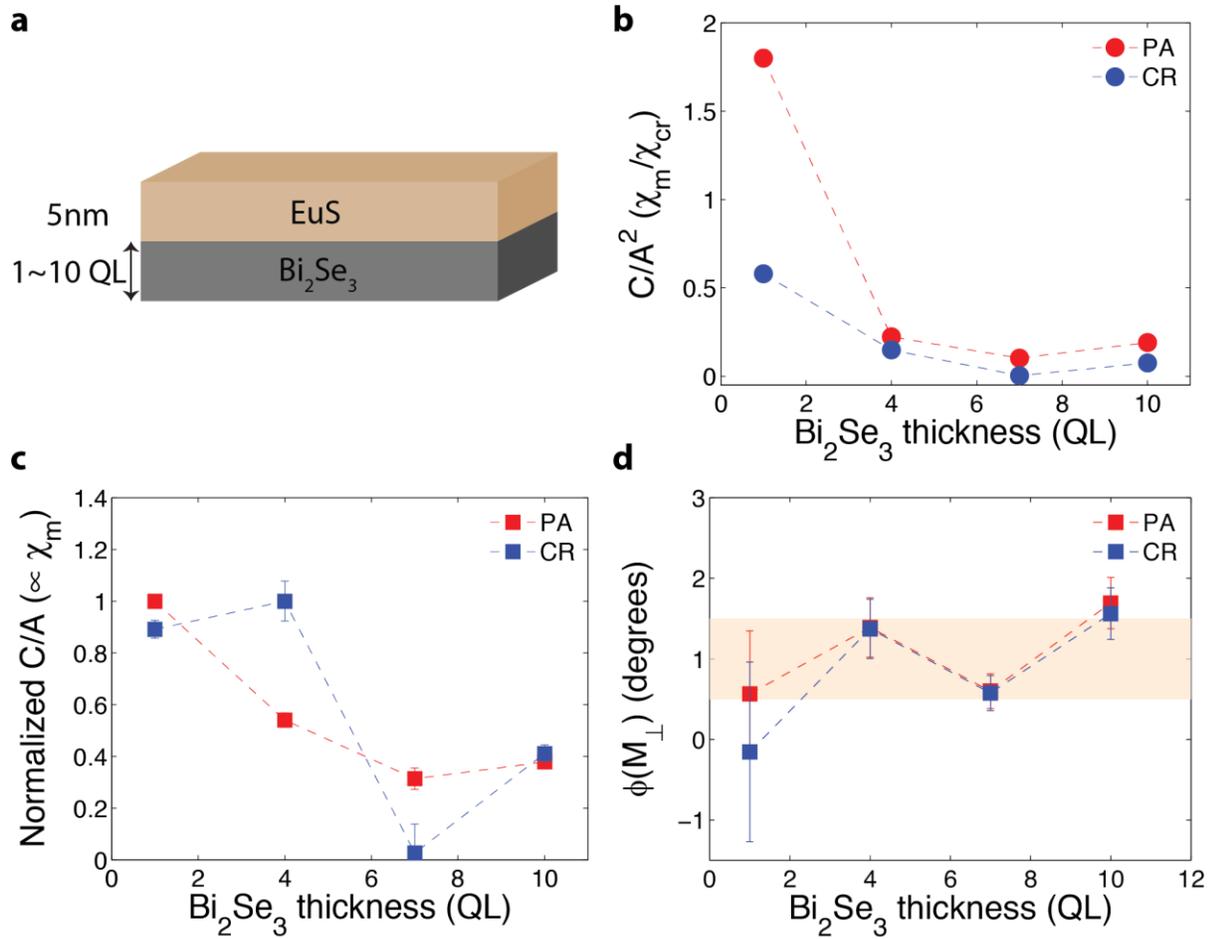

**Figure 4 | Bi₂Se₃ thickness dependence. a,** MSHG parameters were measured from samples of various Bi₂Se₃ thicknesses (1-10 QL). EuS thickness is fixed at 5nm. Both **b,** $C(M)/A^2$ and **c,** $C(M)/A$ show a decreasing behavior, possibly due to increased bulk contributions and SHG absorption from thicker Bi₂Se₃ films. **d**, Out-of-plane response, $\phi(M_\perp)$, for different Bi₂Se₃ thicknesses.

# Supplementary Information

# Direct measurement of proximity-induced magnetism at the buried interface between a topological insulator and a ferromagnet

Changmin Lee[1], Ferhat Katmis[1,2], Pablo Jarillo-Herrero[1], Jagadeesh S. Moodera[1,2] and Nuh Gedik[1*]

[1] Department of Physics, Massachusetts Institute of Technology, Cambridge MA 02139, USA.

[2] Francis Bitter Magnet Laboratory & Plasma Science and Fusion Center, Massachusetts Institute of Technology, Cambridge, MA 02139, USA.

* To whom correspondence and requests for materials should be addressed. Email: gedik@mit.edu

## I. Detailed Experimental Methods

All measurements were carried out while the thin film heterostructure samples were kept under high vacuum ($< 5 \times 10^{-6}$ torr) in an optical cryostat. Incoming laser pulses are focused onto a 50 μm spot using a standard plano-convex lens ($f = 20$cm), and the transmitted laser pulses from the sample are collimated using another convex lens ($f = 12.5$cm).

### (1) SHG Faraday rotation measurements

The collimated pulses are then split into two beams of orthogonal polarizations (*s* and *p*) using a Rochon polarizer. Two photomultiplier tubes (PMTs, Hamamatsu H10721-110) were used to separately detect the SHG intensity of the two polarizations, and a fluorescence filter (Thorlabs MF390-18) were placed before each PMT so that the fundamental wavelength (785 nm) laser pulses are spectrally filtered. Output current of each PMT was terminated with a 10 KΩ resistor, and the voltage across this resistor was then measured using a standard lock-in technique synchronized with the repetition rate of the amplified laser system (30 KHz).

The polarization angle $\phi$ of the output SHG is calculated as follows:

$$\phi = \frac{E_p(2\omega)}{E_s(2\omega)} = \sqrt{\frac{I_p(2\omega)}{I_s(2\omega)}} \tag{S1}$$

where $E_s(2\omega)$, $E_p(2\omega)$, $I_s(2\omega)$ and $I_p(2\omega)$ are the SHG electric fields and intensities of s and p polarizations, respectively. Therefore, the amount of the SHG Faraday rotation angle is

$$\begin{aligned}\Delta\phi &= \phi(+M) - \phi(-M) \\ &= \sqrt{\frac{I_p(2\omega,+M)}{I_s(2\omega,+M)}} - \sqrt{\frac{I_p(2\omega,-M)}{I_s(2\omega,-M)}}\end{aligned} \tag{S2}$$

where $\pm M$ refers to magnetizations of opposite signs.

**(2) MSHG-RA measurements**

In MSHG-RA measurements, a Glan-Taylor polarizer was placed after the collimating beam as an analyzer. The half-wave plate and the polarizer were each mounted on a separate computer-controlled rotation stage (Newport PR50CC), and were simultaneously rotated so that the input and output polarizations are either parallel (PA) or perpendicular (CR, or crossed) to each other. A single PMT was used to take MSHG measurements at each input polarization angle.

**II. Fitting Procedures – MSHG-RA patterns**

**(1) In-plane magnetization**

When an in-plane magnetic field is applied to the sample, the MSHG-RA pattern is described by the following equation (equivalent to equation (3) in the main text) when the input and output polarizations are parallel (PA) to each other[1-4]:

$$I(2\omega) = A^2 \cos^2(3\phi + \phi_1) \pm 2C_{PA}(M) \cos(3\phi + \phi_1) \cos(\phi + \phi_2)$$

$$+ B_{PA}^2(M^2)\cos^2(\phi + \phi_2), \tag{S3}$$

which in turn can be rewritten as:

$$I(2\omega) = \frac{A^2}{2} \cos(6\phi + 2\phi_1) \pm 2C_{PA}(M) \cos(3\phi + \phi_1) \cos(\phi + \phi_2)$$

$$+ \frac{B_{PA}^2}{2}(M^2) \cos(2\phi + 2\phi_2) + D_{PA}. \tag{S4}$$

Similarly, when the input and output polarizations are perpendicular to each other (CR), the MSHG-RA pattern can be described by:

$$I(2\omega) = \frac{A^2}{2} \sin(6\phi + 2\phi_1) \pm 2C_{CR}(M) \sin(3\phi + \phi_1) \sin(\phi + \phi_2)$$

$$+ \frac{B_{CR}^2}{2}(M^2) \sin(2\phi + 2\phi_2) + D_{CR}. \tag{S5}$$

All MSHG-RA data under in-plane magnetic fields are fitted using equations (S4) and (S5), as they minimize the number of products of sinusoidal functions. Four sets of data (two magnetization directions, and two polarization setups) are fitted to two equations sharing nine parameters ($A$, $B_{PA}$, $B_{CR}$, $C_{PA}$, $C_{CR}$, $D_{PA}$, $D_{CR}$, $\phi_1$, $\phi_2$). In all measurements, fitted values of $B$ are too small compared to the size of the error bars (~1-2% of $A$) to extract any information about the strength of the magnetic signal. Data (squares and circles) and fits (lines) from the 7 nm (EuS) – 7 QL ($Bi_2Se_3$) heterostructure sample are shown in Fig. S1.

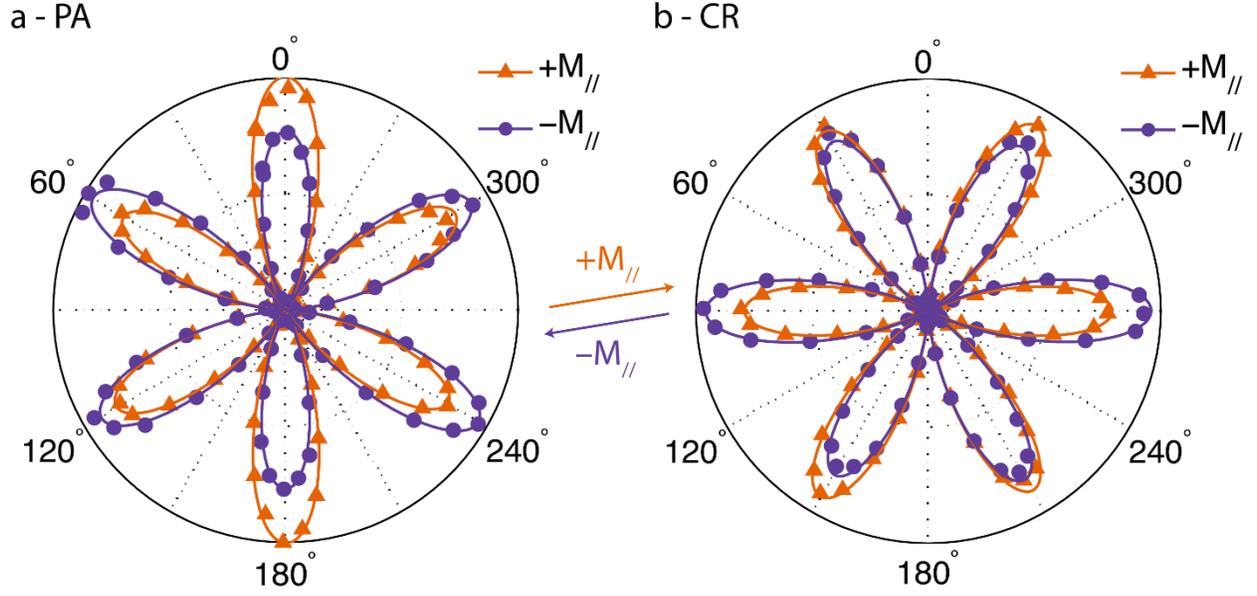

**Figure S1 | MSHG-RA patterns of a 7 nm-7 QL heterostructure film under in-plane magnetic fields.** MSHG-RA patterns are shown for the **a,** PA and **b,** CR polarization setups. Magnetic field are denoted by the orange and purple arrows.

**(2) Out-of-plane magnetization**

Similarly, MSHG data under an out-of-plane magnetic field are fitted using the following two equations, depending on the polarization combinations (PA and CR),

$$PA: I(2\omega) = \frac{A^2}{2}\cos(6\phi + 2\phi_1 \pm \phi_{PA}(M_\perp))$$
$$\pm 2C_{PA}(M)\cos(3\phi + \phi_1 \pm \phi_{PA}(M_\perp))\cos(\phi + \phi_2) \quad (S6)$$
$$+ \frac{B_{PA}^2}{2}(M^2)\cos(2\phi + 2\phi_2) + D_{PA}$$

$$CR: I(2\omega) = \frac{A^2}{2}\cos(6\phi + 2\phi_1 \pm \phi_{CR}(M_\perp))$$

$$\pm 2C_{PA}(M)\cos(3\phi + \phi_1 \pm \phi_{CR}(M_\perp))\cos(\phi + \phi_2) \quad (S7)$$

$$+ \frac{B_{CR}^2}{2}(M^2)\cos(2\phi + 2\phi_2) + D_{CR}.$$

As described in the main text, a tilted (~4°) magnetic field was applied to the sample so that both the in-plane and out-of-plane magnetic moments are aligned along a preferred direction. The magnetic field strengths are ± 4000 Oe along the out-of-plane, and ± 400 Oe along the in-plane direction. We obtained a total of 8 sets of data (4 magnetic field directions and 2 polarization setups), which are fitted using equations (S6) and (S7). Data and fits for the 7nm – 7QL heterostructure sample are shown in Fig. S2.

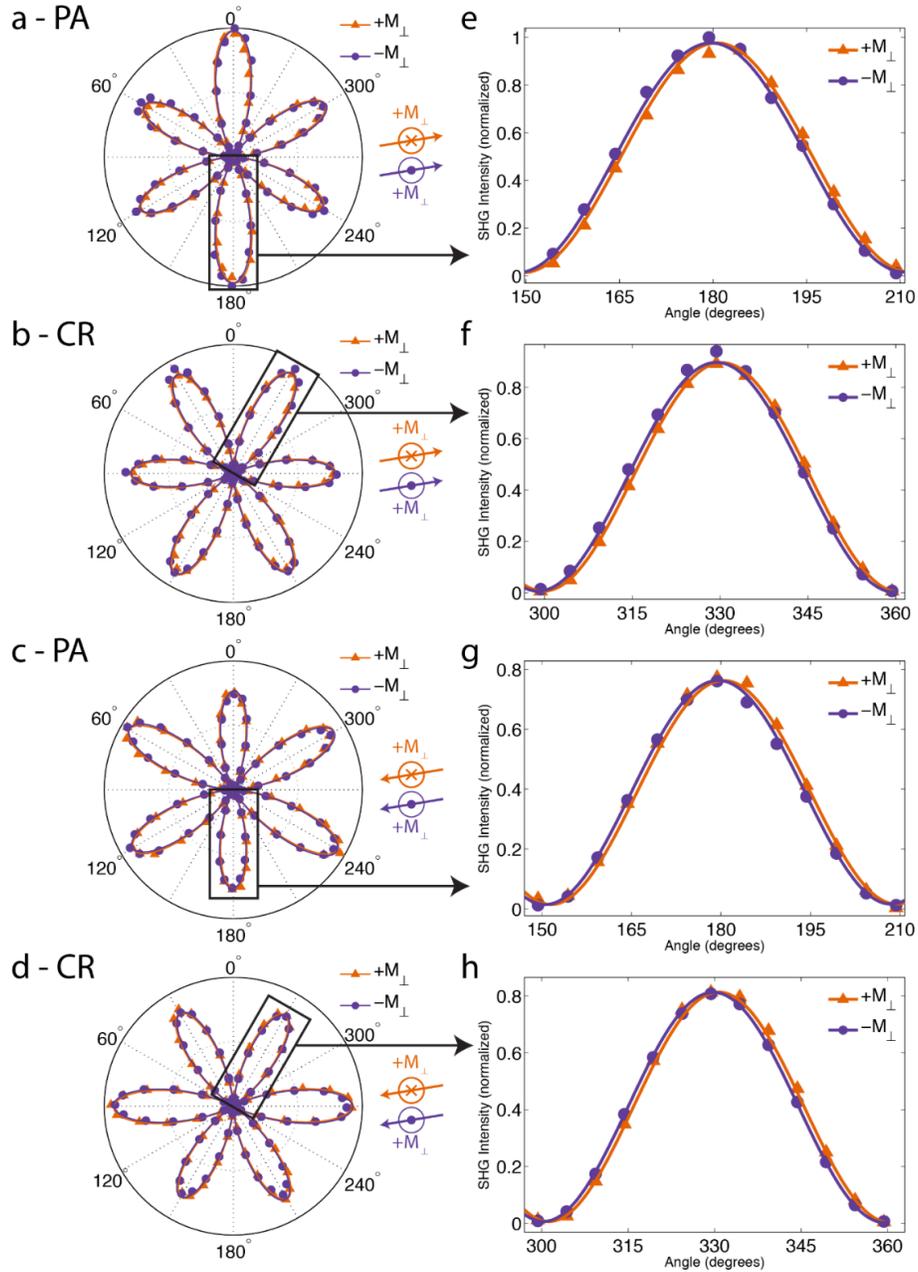

**Figure S2 | MSHG-RA patterns of a 7 nm-7 QL heterostructure film under out-of-plane magnetic fields. a-d,** MSHG-RA patterns are shown for the PA (**a,c**) and CR (**b,d**) polarization setups while a canted magnetic field (∼4°) was applied to the sample. Canted magnetization directions are denoted by arrows, circles, and crosses next to each figure. **e-h,** enlarged plots of the squared area in **a-d**.

### III. Quantum well states (QWS) and magnetic second harmonic generation (MSHG)

When the thickness of a material becomes finite (say, along the z direction), as in the case of the EuS-$Bi_2Se_3$ heterostructures, electrons can form a QWS bound along the z direction. The energy dispersion along the z axis now becomes more discretized. As the thickness of the sample is varied, the number of discrete points in the band structure also changes, and the $k_z$ points are shifted accordingly (Fig. S3). SHG is a two-photon process that involves a resonant transition from one state in the occupied band to another state in the unoccupied band. Upon changing the thickness of the quantum well, optical transition takes place at different $k_z$ values, leading to a change in the SHG magnitude.

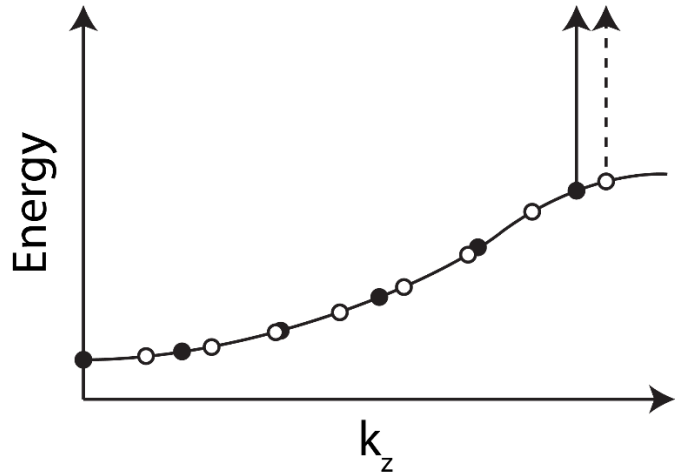

**Figure S3 | Band structure of a model QWS**. In two QWSs of different thicknesses, optical transitions take place at different points of the Brillouin zone. Filled and empty circles correspond to the $k_z$ points of the thicker and the thinner QWS, respectively.

A more subtle effect on MSHG can be found in thin film heterostructures composed of both ferromagnetic and paramagnetic materials. In the past, large variations in the MSHG signal were measured in Co/Cu[5] and Au/Co/Au[6] heterostructures of different thicknesses. In both cases, the MSHG intensity shows a strong oscillatory behavior as a function of the paramagnetic layer thickness. In the former case, the minority spin states of Co have a similar density of states (DOS) at the Fermi level as that of Cu, whereas the majority spin states have a very different DOS. Therefore, the quantum well confinement strength of the two spin states are different from each other. As the thickness of the paramagnetic Cu layer is tuned, the DOS at the Fermi level of Cu correspondingly changes, and it in fact oscillates as a function of thickness[7]. Therefore, the MSHG response in turn exhibits an oscillatory behavior as a function of the thickness of the paramagnetic layer in both the Co/Cu[5] and Au/Co/Au[6] heterostructures. A similar argument can be applied to the strong fluctuating MSHG response from the $EuS$-$Bi_2Se_3$ heterostructures. Since EuS is a good electrical insulator, it is unlikely that a QWS resides inside the bulk of EuS, as most electrons are confined within the nuclei[8]. In regard to the QWS inside the $Bi_2Se_3$ film, the majority and minority spin states have different DOS at the Fermi level when the $Bi_2Se_3$ film is magnetized through the proximity effect from the neighboring EuS film. Again, the strength of the potential barrier exerted to each spin state is different from each other. As the thickness of the EuS film is tuned, a change in the band structure of the EuS film can cause the relative confinement strength of the two spin states to become different, resulting in a change in the MSHG response.